\documentclass[twocolumn,10pt]{article}
\usepackage{amsmath,amssymb}
\usepackage{graphicx}

\makeatletter \oddsidemargin-.25in \evensidemargin-.25in
\makeatother

\begin{document}

\date{}
\title{\Large\bf Solving Stochastic Differential Equations with
Jump-Diffusion Efficiently: Applications to FPT Problems in Credit
Risk}
\author{Di Zhang$^1$ and Roderick V.N. Melnik$^1$\\[2mm]
 $^1$Mathematical Modelling and Computational Sciences, Wilfrid
Laurier University,\\ Waterloo, ON, Canada N2L 3C5, Email:
rmelnik@wlu.ca} \maketitle

\pagestyle{empty}

{\footnotesize \noindent {\bf Abstract.}  The first passage time
(FPT) problem is ubiquitous in many applications. In finance, we
often have to deal with stochastic processes with jump-diffusion, so
that the FTP problem is reducible to a stochastic differential
equation with jump-diffusion. While the application of the
conventional Monte-Carlo procedure is possible for the solution of
the resulting model, it becomes computationally inefficient which
severely restricts its applicability in many practically interesting
cases. In this contribution, we focus on the development of
efficient Monte-Carlo-based computational procedures for solving the
FPT problem under the multivariate (and correlated) jump-diffusion
processes. We also discuss the implementation of the developed
Monte-Carlo-based technique for multivariate jump-diffusion
processes driving by several compound Poisson shocks. Finally, we
demonstrate the application of the developed methodologies for
analyzing the default rates and default correlations of differently
rated firms
via historical data.
}
\vskip.2in

\section{Introduction}
\label{introduction}

\noindent Many problems in finance require the information on the
first passage time (FPT) of a stochastic process. Mathematically,
such problems are often reduced to the evaluation of the probability
density of the time for a process to cross a certain level. Recent
research in finance theory has renewed the interest in
jump-diffusion processes (JDP), and the FPT problem for such
processes is applicable to several finance problems, such as pricing
barrier options \cite{Metwally:2002,Zhang:1997}, credit risk
analysis \cite{Zhou:2001:jump}. While in other areas of applications
the FPT problem can often be solved analytically, in finance we
usually have to resort to the application of numerical procedures,
in particular when we deal with jump-diffusion processes.

Among numerical procedures, Monte-Carlo methods remain a primary
candidate for applications. However, the conventional Monte-Carlo
procedure becomes computationally inefficient when it is applied to
the jump-diffusion processes. Many researchers have contributed to
the field of enhancement of the efficiency of Monte-Carlo
simulations. Atiya and Metwally \cite{Metwally:2002,Atiya:2005} have
recently developed a fast Monte-Carlo-type numerical method to solve
the FPT problem in the one-dimensional case.

Note that apart from the pricing and hedging of options on a single
asset, practically all financial applications require a multivariate
model with dependence between different assets. In particularly,
jumps in the price process must be taken into account in most of the
applications \cite{Cont:2003,Crescenzo:2005}. In this contribution,
we generalize our previous fast Monte-Carlo method (for
non-correlated jump-diffusion cases) to multivariate (and
correlated) jump-diffusion processes. The developed technique
provides an efficient tool for a number of applications, including
credit risk and option pricing. We also discuss the implementation
of the developed Monte-Carlo-based techniques for a subclass of
multidimensional L\'{e}vy processes with several compound Poisson
shocks. Finally, we demonstrate the applicability of this technique
to the analysis of the default rates and default correlations of two
different correlated firms via a set of empirical data.

The article is organized as follows: Section \ref{model} describes
the mathematical model. The methodologies are presented in Section
\ref{methodology}, applications and discussions are given in Section
\ref{application}. Concluding remarks are given in Section
\ref{conclusion}.

\section{Mathematical model}
\label{model}

\noindent In this section, first we present a probabilistic
description of default events and default correlations in finance.
Next, we describe the multivariate jump-diffusion processes and
provide details on the first passage time distribution under the
one-dimensional Brownian bridge (the algorithms which is used to
generate correlated first passage times will be described in Section
\ref{methodology}). Finally, we present kernel estimation in the
context of our problem that can be used to represent the first
passage time density function.

In finance, a firm $i$ defaults when it can not meet its financial
obligations, or in other words, when the firm assets value $V_i(t)$
falls below a threshold level $D_{V_i}(t)$. In this contribution, we
use an exponential form defining the threshold level
$D_{V_i}(t)=\kappa_i\exp(\gamma_i t)$ as proposed by Black and Cox
\cite{Black-Cox:1976}, where $\gamma_i$ can be interpreted as the
growth rate of firm's liabilities. Coefficient $\kappa_i$ captures
the liability structure of the firm and is usually defined as a
firm's short-term liability plus 50\% of the firm's long-term
liability \cite{Zhou:2001:corr}. If we set $X_i(t)=\ln[V_i(t)]$,
then the threshold of $X_i(t)$ is $D_i(t)=\gamma_i t+\ln(\kappa_i)$.
Our main interest is in the process $X_i(t)$.

In the market economy, individual companies are inevitably linked
together via dynamically changing economic conditions
\cite{Zhou:2001:corr}. Therefore, the default events of companies
are often correlated, especially in the same industry. Take two
firms $i$ and $j$ as an example, whose probabilities of default are
$P_i$ and $P_j$, respectively. Then the default correlation can be
defined as
\begin{equation}
  \rho_{ij}=\frac{P_{ij}-P_{i}P_{j}}{\sqrt{P_i(1-P_i)P_j(1-P_j)}},
  \label{Eq:corr}
\end{equation}
where $P_{ij}$ is the probability of joint default.

Zhou \cite{Zhou:2001:corr} and Hull et al \cite{Hull:2001} were the
first to incorporate default correlation into the Black-Cox first
passage structural model. They have obtained the closed form
solutions for the joint probability of firm 1 to default before
$T_1$ and firm 2 to default before $T_2$. However, none of the above
known models includes jumps in the processes. At the same time, it
is well-known that jumps are a major factor in the credit risk
analysis. With jumps included in such analysis, a firm can default
instantaneously because of a sudden drop in its value which is
impossible under a diffusion process \cite{Zhou:2001:jump}.
Therefore, for multiple processes, considering the simultaneous
jumps can be a better way to estimate the correlated default rates.

A natural approach to introduce jumps into a multidimensional model
is to utilize the compound Poisson shocks. The dates of market
crashes can be modeled as arrival times of a standard Poisson
process $N_t$, which leads us to the following model for the
log-price processes of $d$ assets \cite{Cont:2003}:
\begin{equation}
  \left\{\begin{array}{l}
    X_i(t)=\mu_i t+B_i(t)+Z_i(t),\;i=1,2,\cdots,d,\\
    Z_i(t)=\sum_{j=1}^{N_t}Y_{ij},
  \end{array}\right.
  \label{equation:MJDP:one}
\end{equation}
where $B(t)$ is a $d$-dimensional Brownian motion with covariance
matrix $\sigma=(\sigma_{ij})$ which can be written as
\[
  B_i(t)=\sum_{j=1}^{d}\sigma_{ij}W_j(t),
\]
and $W_j(t)$ is the standard Brownian motion. For $i$-th asset,
$\{Y_{ij}\}_{j=1}^\infty$ are i.i.d. $d\mathrm{-dimensional}$ random
vectors which determine the sizes of jumps in individual assets
during a market crash. At the $j$-th shock, the jump-sizes of
different assets $Y_{ij}$ may be correlated.

This model contains only one driving Poisson shock which stands for
that the global market crash affecting all assets. Sometimes it is
necessary to have several independent shocks to account for events
that affect individual companies or individual sectors rather than
the entire market. In this case we need to introduce several driving
Poisson processes into the model, which now takes the following form
\cite{Cont:2003}:
\begin{equation}
  X_i(t)=\mu_i t+B_i(t)+\sum_{k=1}^{m}\sum_{j=1}^{N_t^k}Y_{ijk},\;i=1,2,\cdots,d,\\
  \label{equation:MJDP:several}
\end{equation}
where $N_t^1$, $\cdots$, $N_t^m$ are Poisson processes driving $m$
independent shocks and $Y_{ijk}$ is the size of jump in $i$-th
component after $j$-th shock of type $k$. The vectors
$\{Y_{ijk}\}_{i=1}^d$ for different $j$ and/or $k$ are independent.

First, let us consider a firm $i$, as described by Eq.
(\ref{equation:MJDP:one}), such that its state vector $X_i$
satisfies the following stochastic differential equation:
\begin{eqnarray}
  dX_i & = & \mu_{i}dt+\sum_{j=1}^{d}\sigma_{ij}dW_j+dZ_i\nonumber\\
       & = & \mu_{i}dt+\sigma_{i}dW_i+dZ_i,
  \label{JDP:one}
\end{eqnarray}
where $W_i$ is a standard Brownian motion and
$\sigma_{i}=\sqrt{\sum_{j=1}^{d}\sigma_{ij}^2}$.

We assume that in the interval $[0,T]$, the total number of jumps
for firm $i$ is $M_i$. Let the jump instants be $T_{1},
T_{2},\cdots,T_{M_i}$. Let $T_{0}=0$ and $T_{M_i+1}=T$. The
quantities $\tau_j$ equal to interjump times, which are
$T_{j}-T_{j-1}$. Following the notation of \cite{Atiya:2005}, let
$X_{i}(T_{j}^{-})$ be the process value immediately before the $j$th
jump, and $X_{i}(T_{j}^{+})$ be the process value immediately after
the $j$th jump. The jump-size is $X_{i}(T_{j}^{+})-X_{i}(T_{j}^{-})$
and we can use such jump-sizes to generate $X_{i}(T_{j}^{+})$
sequentially.

Although for jump-diffusion processes, the closed form solutions are
usually unavailable, yet between each two jumps the process is a
Brownian bridge for one-dimensional jump-diffusion process. Let
$B(s)$ be a Brownian bridge in the interval $[T_{j-1},T_{j}]$ with
$B(T_{j-1}^{+})=X_i(T_{j-1}^{+})$ and $B(T_{j}^{-})=X_i(T_{j}^{-})$.
If $X_i(T_{j}^{-})>D_{i}(t)$, then the probability that the minimum
of $B(s_i)$ is always above the boundary level is \cite{Atiya:2005}
\begin{eqnarray}
  P_{ij} =
  1-e^{-\frac{2[X_i(T_{j-1}^{+})-D_{i}(t)][X_i(T_{j}^{-})-D_{i}(t)]}{\tau_{j}\sigma_{i}^{2}}}.
   \label{BM:default}
\end{eqnarray}
This implies that $B(s_i)$ is below the threshold level, which means
the default happens or already happened, and its probability is
$1-P_{ij}$.

For firm $i$, after generating a series of first passage times
$s_i$, we use a kernel density estimator with Gaussian kernel to
estimate the first passage time density (FPTD) $f$. The kernel
density estimator is based on centering a kernel function of a
bandwidth as follows:
\begin{equation}
  \widehat{f}=\frac{1}{N}\sum_{i=1}^{N}K(h,t-s_{i}).
  \label{Eq:estimator}
\end{equation}

\noindent The optimal bandwidth in the kernel function $K$ can be
calculated as \cite{Silverman:1986}:
\begin{equation}
  h_{opt}=\left(2N\sqrt{\pi}\int_{-\infty}^{\infty}(f_{t}'')^{2}dt\right)^{-0.2},
  \label{estamate:hopt}
\end{equation}
where $N$ is the number of generated points and $f_{t}$ is the true
density. Here we use the approximation for the distribution as a
gamma distribution as proposed in \cite{Atiya:2005}:
\begin{equation}
  f_{t}=\frac{\alpha^{\beta}}{\Gamma(\beta)}t^{\beta-1}\exp(-\alpha t).
\end{equation}

\section{Methodology of the solution}
\label{methodology}

\begin{figure}[hbtp]
  \centering
  \includegraphics[width=7.5cm]{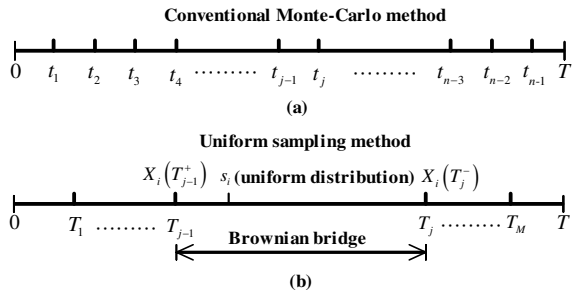}
  \caption[Schematic diagram of the conventional Monte-Carlo and
   uniform sampling method]
  {Schematic diagram of (a) the conventional Monte-Carlo and
   (b) the uniform sampling (UNIF) method.}
  \label{Fig:Method}
\end{figure}

\noindent Let us recall the conventional Monte-Carlo procedure in
application to the analysis of the evolution of firm $X_i$ within
the time horizon $[0,T]$. We divide the time horizon into $n$ small
intervals $[0,t_1]$, $[t_1,t_2]$, $\cdots$, $[t_{n-1},T]$ as shown
in Fig. \ref{Fig:Method}(a). In each Monte-Carlo run, we need to
calculate the value of $X_i$ at each discretized time $t$ without
explicitly distinguishing the effects of the jump and diffusion
terms \cite{Glasserman:2004}. As usual, in order to reduce
discretization bias, the number $n$ must be large
\cite{Platen:2003}.
The conventional Monte-Carlo procedure exhibits substantial
computational difficulties when applied to jump-diffusion processes.
Indeed, for a typical jump-diffusion process, as shown in Fig.
\ref{Fig:Method}(b), let $T_{j-1}$ and $T_j$ be any successive jump
instants, as described above. Then, in the conventional Monte-Carlo
method, although there is no jump occurring in the interval
$[T_{j-1},T_j]$, yet we need to evaluate $X_i$ at each discretized
time $t$ in $[T_{j-1},T_j]$. This very time-consuming procedure
results in a serious shortcoming of the conventional Monte-Carlo
methodology.

To remedy the situation, two modifications of the conventional
procedure were recently proposed \cite{Metwally:2002,Atiya:2005}
that allow us a potential speed-up of the conventional methodology
in 10-30 times. One of the modifications, the uniform sampling
method (UNIF), involves sampling using uniform distribution. The
other, inverse Gaussian density sampling is based on the inverse
Gaussian density method for sampling. Both methodologies were
developed for the univariate case.

The major improvement of the uniform sampling method is based on the
fact that it only evaluates $X_i$ at generated jump times, while
between each two jumps the process is a Brownian bridge (see Fig.
\ref{Fig:Method}(b)). Hence, we just consider the probability of
$X_i$ crossing the threshold in $(T_{j-1},T_j)$ instead of
evaluating $X_i$ at each discretized time $t$. More precisely, in
the uniform sampling method, we assume that the values of
$X_i(T_{j-1}^+)$ and $X_i(T_j^-)$ are known as two end points of the
Brownian bridge, the probability of firm $i$ defaults in
$(T_{j-1},T_j)$ is $1-P_{ij}$ which can be computed according to Eq.
(\ref{BM:default}). Then we generate a variable $s_i$ from a
distribution uniform in an interval
$[T_{j-1},T_{j-1}+\frac{T_{j}-T_{j-1}}{1-P_{ij}}]$. If the generated
point $s_i$ falls in the interjump interval $[T_{j-1},T_j]$, then we
have successfully generated a first passage time $s_i$ and can
neglect the other intervals and perform another Monte-Carlo run. On
the other hand, if the generated point $s_i$ falls outside the
interval $[T_{j-1},T_j]$ (which happens with probability $P_{ij}$),
then that point is ``rejected''. This means no boundary crossing has
occurred in the interval, and we proceed to the next interval and
repeat the whole process again.

In what follows, we focus on the further development of the uniform
sampling method and extend it to multivariate and correlated
jump-diffusion processes. First, we consider there is only one
driving Poisson shock with arrival rate $\lambda$ as described in
Eq. (\ref{equation:MJDP:one}). The distribution of $(T_j-T_{j-1})$
are the same for each firm. The jump-size can be generated by a
given distribution which can be different for different firms to
reflect specifics of the jump process for each firm. In order to
implement the UNIF method for the multivariate processes, we need to
consider several points:
\begin{enumerate}
  \item We exemplify our description by considering an exponential
distribution (mean value $\mu_T$) for $(T_j-T_{j-1})$ and a normal
distribution (mean value $\mu_J$ and standard deviation $\sigma_J$)
for the jump-size. We can use any other distribution when
appropriate.
  \item An array \texttt{IsDefault} (whose size is
the number of firms denoted by $N_{\mathrm{firm}}$) is used to
indicate whether firm $i$ has defaulted in this Monte-Carlo run. If
the firm defaults, then we set \texttt{IsDefault}$(i)=1$, and will
not evaluate it during this Monte-Carlo run.
  \item Most importantly, as we have mentioned before, the default events of
firm $i$ are inevitably correlated with other firms, for example
firm $i+1$. Hence, firm $i$'s first passage time $s_i$ is indeed
correlated with $s_{i+1}$ -- the first passage time of firm $i+1$.
We must generate several correlated $s_i$ in each interval
$[T_{j-1},T_{j-1}+\frac{T_{j}-T_{j-1}}{1-P_{ij}}]$ which is the key
point for multivariate correlated processes.

Note that each process is a Brownian motion in the interval
$[T_{j-1},T_{j}]$, so we can compute the correlation coefficient
$\rho_{i,i+1}$ of firms $i$ and $i+1$ by using Zhou's model without
jumps \cite{Zhou:2001:corr} and then use this value for modeling
correlated $s_i$ and $s_{i+1}$. Let us introduce a new variable
$b_{ij}=\frac{T_{j}-T_{j-1}}{1-P_{ij}}$. Then we have
$s_i=b_{ij}Y_i+T_{j-1}$, where $Y_i$ are uniformly distributed in
$[0,1]$. Moreover, the correlation of $Y_i$ and $Y_{i+1}$ is the
same as $s_i$ and $s_{i+1}$. The correlated uniform random variables
$Y_1, Y_2,\cdots$ can be generated by using the sum-of-uniforms
(SOU) method \cite{Chen:2005}.
\end{enumerate}

Next, we will describe our algorithm for multivariate jump-diffusion
processes, which is an extension of the one-dimensional case
developed earlier by other authors (e.g.
\cite{Metwally:2002,Atiya:2005}).

Consider $N_{\mathrm{firm}}$ firms in the given time horizon
$[0,T]$. First, we generate the jump instant $T_{j}$ by generating
interjump times $(T_{j}-T_{j-1})$ and set all the
\texttt{IsDefault}$(i)=0$ to indicate that no firm defaults at
first.

From Fig. \ref{Fig:Method}(b) and Eq. (\ref{JDP:one}), we can
conclude that for each process $X_i$ we can make the following
observations:
\begin{enumerate}
  \item If no jump occurs, as described by Eq. (\ref{JDP:one}), the
interjump size $(X_i(T_{j}^{-})-X_i(T_{j-1}^{+}))$ follows a normal
distribution of mean $\mu_i(T_{j}-T_{j-1})$ and standard deviation
$\sigma_i\sqrt{T_{j}-T_{j-1}}$. We get
\[
X_i(T_{j}^{-})\sim X_i(T_{j-1}^{+})+\mu_i\tau_{j}+
\sigma_{i}N(0,\tau_{j}),
\]
where the initial state is $X_i(0)=X_i(T_{0}^{+})$.
  \item If jump occurs, we simulate the jump-size by a normal distribution
  or another distribution when appropriate, and compute the postjump value:
\[
  X_i(T_{j}^{+})=X_i(T_{j}^{-})+Z_i(T_{j}).
\]
\end{enumerate}

This completes the procedure for generating beforejump and postjump
values $X_i(T_{j}^{-})$ and $X_i(T_{j}^{+})$. As before,
$j=1,\cdots,M$ where $M$ is the total number of jumps for all the
firms. We compute $P_{ij}$ according to Eq. (\ref{BM:default}). To
recur the first passage time density (FPTD) $f_i(t)$, we have to
consider three possible cases that may occur for each non-default
firm $i$ (for more details, we refer to references
\cite{Zhang:2006a} and \cite{Zhang:2006b}):
\begin{enumerate}
  \item \textbf{First passage happens inside the interval.} We know that if
$X_i(T_{j-1}^{+})>D_i(T_{j-1})$ and $X_i(T_{j}^{-})<D_i(T_{j})$,
then the first passage happened in the time interval
$[T_{j-1},T_{j}]$. To evaluate when the first passage happened, we
introduce a new viable $b_{ij}$ as
$b_{ij}=\frac{T_{j}-T_{j-1}}{1-P_{ij}}$. We generate several
correlated uniform numbers $Y_i$ by using the SOU method, then
compute $s_i=b_{ij}Y_i+T_{j-1}$. If $s_i$ belongs to interval
$[T_{j-1},T_{j}]$, then the first passage time occurred in this
interval. We set \texttt{IsDefault}$(i)=1$ to indicate firm $i$ has
defaulted. To get the density for the entire interval $[0,T]$, we
use
$\widehat{f}_{i,n}(t)=\left(\frac{T_{j}-T_{j-1}}{1-P_{ij}}\right)g_{ij}(s_i)*K(h_{opt},t-s_i)$,
where $n$ is the iteration number of the Monte-Carlo cycle, and
$g_{ij}(s_i)$ is the conditional boundary crossing density used to
obtain an appropriate density estimate \cite{Atiya:2005}.
  \item \textbf{First passage does not happen in this interval.}
If $s_i$ does not belong to interval $[T_{j-1},T_{j}]$, then the
first passage time has not yet occurred in this interval.
  \item \textbf{First passage happens at the right boundary of the interval.} If
$X_i(T_{j}^{+})<D_i(T_{j})$ and $X_i(T_{j}^{-})>D_i(T_{j})$, then
$T_{j}$ is the first passage time. We evaluate the density function
using kernel function $\widehat{f}_{i,n}(t)=K(h_{opt},t-T_{j})$, and
set \texttt{IsDefault}$(i)=1$.
\end{enumerate}

Next, we increase $j$ and examine the next interval and analyze the
above three cases for each non-default firm again. After running $N$
times Monte-Carlo cycle, we get the FPTD of firm $i$ as
$\widehat{f}_{i}(t)=\frac{1}{N}\sum_{n=1}^{N}\widehat{f}_{i,n}(t)$.

\section{Generalizations}
\label{generalizations}

\noindent In the above algorithms, we only consider one driving
Poisson shock that affecting all the firms. Sometimes it is
necessary to have several independent shocks to account for events
that affect individual companies rather than the entire market.
Hence, our next goal is to generalize the developed multivariate
uniform sampling (MUNIF) method to the case of several independent
Poisson shocks.

We consider $d$ firms which are driven by $m$ independent Poisson
shocks $N_t^1$, $\cdots$, $N_t^m$ as described by Eq.
(\ref{equation:MJDP:several}). Let $M_k$ be the number of jumps for
each Poisson shock $N_t^k$, $M=\sum_{k=1}^{m}M_k$ is the total
number of jumps. We generate the jump instant $T_{j,k}$ by
generating the interjump times $(T_{j,k}-T_{j-1,k})$ for each
Poisson shock. Then we sort the jump instant $T_{j,k}$ by the
relevant ascending order and still denote them as
$T_j\;(j=1,2,\cdots,M)$. Furthermore, an array \texttt{ShockType}
(whose size is $M$) is used to record the type of shock at $T_j$.
Then we can carry out multivariate uniform sampling method for this
case as same as in Section \ref{methodology}, but the postjump value
should be calculated as:
\[
  X_i(T_{j}^{+})=X_i(T_{j}^{-})+Y_i(T_{j}),
\]
where $Y_i(T_{j})$ is the size of jump for $i$-th firm at $T_j$, the
type of shock is determined by the array \texttt{ShockType}.
Besides, we may generate correlated $Y_i(T_{j})$ for different
firms.

\section{Applications and discussions}
\label{application}

In this section, we demonstrate the developed model at work for
analyzing the default events of two differently rated and correlated
firms (denoted as A- and Ba-rated firms according to the Moody's
debt rating system) via a set of historical default data as
presented by \cite{Zhou:2001:corr}. Our first task is to describe
the first passage time density functions and default rates of these
firms.

In order to apply our developed procedure, first we need to
calibrate the developed model, in other words, to numerically choose
or optimize the parameters, such as drift, volatility and jumps to
fit the most liquid market data. As mentioned in Section
\ref{methodology}, after Monte-Carlo simulation we obtain the
estimated density $\widehat{f}_{i}(t)$ by using the kernel estimator
method. The cumulative default rates for firm $i$ in our model is
defined as,
\begin{equation}
  P_i(t)=\int_{0}^{t}\widehat{f}_{i}(\tau)d\tau.
\end{equation}
Then we minimize the difference between our model and historical
default data $\widetilde{A}_i(t)$ to obtain the optimized parameters
in the model (such as $\sigma_{ij}$, arrival intensity $\lambda$ in
Eq. (\ref{JDP:one})):
\begin{equation}
  \mathrm{argmin}\left(\sum_{i}\sqrt{\sum_{t_j}\left(
    \frac{P_{i}(t_j)-\widetilde{A}_i(t_j)}{t_j}\right)^2}\right).
  \label{Eq:calibration:default}
\end{equation}

\noindent For convenience, we reduce the number of optimizing
parameters by:
\begin{enumerate}
  \item Setting $X(0)=2$ and $\ln(\kappa)=0$.
  \item Setting the growth rate $\gamma$ of debt value equivalent to the
growth rate $\mu$ of the firm's value \cite{Zhou:2001:corr}, so the
default of firm is non-sensitive to $\mu$. In our computations, we
set $\mu=-0.001$.
  \item The interjump times $(T_j-T_{j-1})$ satisfy an exponential
  distribution with mean value equals to 1.
  \item The arrival rate for jumps satisfies the Poisson distribution with
intensity parameter $\lambda$, where the jump size is a normal
distribution $Z_t\sim N(\mu_{Z},\sigma_{Z})$.
\end{enumerate}

As a result, we only need to optimize $\sigma$, $\lambda$,
$\mu_{Z}$, $\sigma_{Z}$ for each firm. This is done by minimizing
the difference between our simulated default rates and historical
data. Moreover, we assume there is only one driving Poisson shock
with arrival rate $\lambda$. Hence we first optimize four parameters
for, e.g., the A-rated firm, and then set the same $\lambda$ for
Ba-rated firm.

The minimization was performed by the using quasi-Newton procedure
implemented as a Scilab program. The optimized parameters are
described in Table \ref{Table:param:one}.
\begin{table}[htbp]
  \centering
  \caption{Optimized parameters for A- and Ba-rated firms by using the MUNIF
method. In each step of the optimization, we choose the Monte Carlo
runs $N=50,000$.}
  \label{Table:param:one}
  \begin{tabular}{l|cccc}
    \hline
    & $\sigma$ & $\lambda$ & $\mu_{Z}$ & $\sigma_{Z}$\\
    \hline
    A   & 0.0900 & 0.1000 & -0.2000 & 0.5000 \\
    Ba  & 0.1587 & 0.1000 & -0.5515 & 1.6412 \\
    \hline
  \end{tabular}
\end{table}

\begin{figure}[hbtp]
  \centering
  \includegraphics[width=6.8cm]{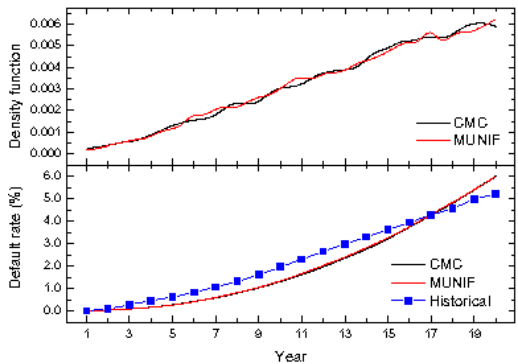}
  \caption{Density function (top) and default rate (bottom) of A-rated firm.
The simulations were performed with Monte-Carlo runs $N=100,000$,
for the conventional Monte-Carlo method, the discretization size of
time horizon was $\Delta=0.005$.}
  \label{Fig:FirmA}
\end{figure}

\begin{figure}[hbtp]
  \centering
  \includegraphics[width=6.8cm]{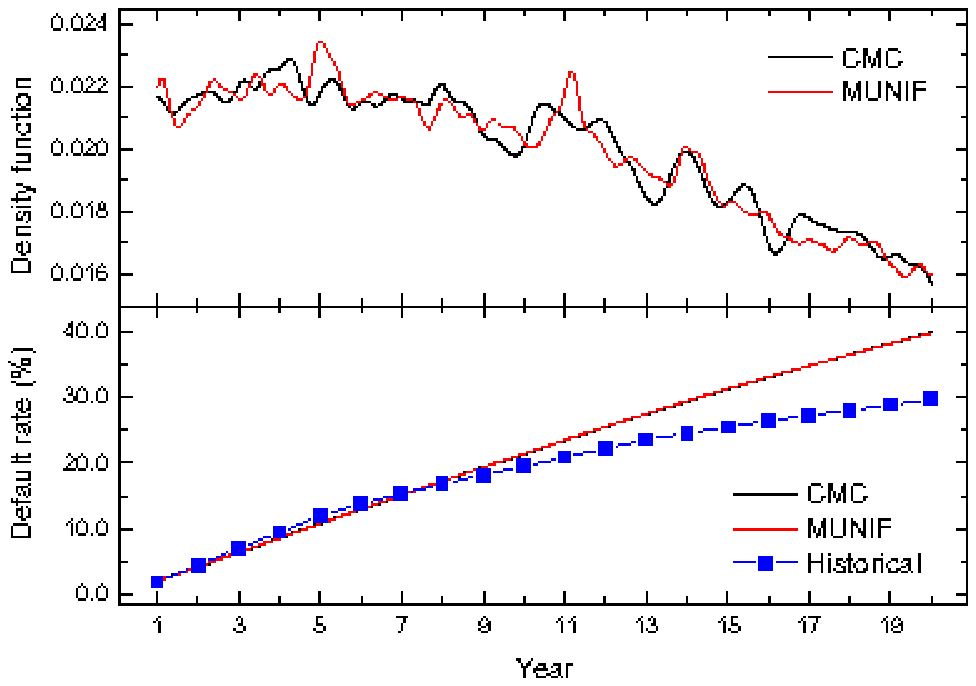}
  \caption{Density function (top) and default rate (bottom) of Ba-rated firm.
The simulations were performed with Monte-Carlo runs $N=100,000$,
for the conventional Monte-Carlo method, the discretization size of
time horizon was $\Delta=0.005$.}
  \label{Fig:FirmBa}
\end{figure}

\noindent By using these optimized parameters, we carried out the
final simulation with Monte Carlo runs $N=100,000$. Moreover, we
also carried out simulations based on the conventional Monte-Carlo
method with the same parameters and the discretization size of time
horizon $\Delta=0.005$. The estimated first passage time density
functions of A- and Ba-rated firms are shown in the top of Fig.
\ref{Fig:FirmA} and \ref{Fig:FirmBa}, respectively. The simulated
cumulative default rates (line) together with historical data
(squares) are given in the bottom of Fig. \ref{Fig:FirmA} and
\ref{Fig:FirmBa}. In Table \ref{Tab:hopt:CPU}, we give the
calculated optimal bandwidths and the corresponding CPU times.
\begin{table}[hbtp]
  \centering
  \caption{The optimal bandwidth $h_{opt}$ and CPU time per Monte-Carlo run
of the simulations. The simulations were performed with Monte-Carlo
runs $N=100,000$, for the conventional Monte-Carlo (CMC) method, the
discretization size of time horizon was $\Delta=0.005$.}
  \label{Tab:hopt:CPU}
  \begin{tabular}{l|lcc}
    \hline
    & & $h_{opt}$ & CPU time\\
    \hline
    A & CMC  & 0.892333 & 0.119668\\
      & UNIF & 0.653902 & 0.000621\\
    \hline
    Ba & CMC  & 0.427252 & 0.119675\\
       & UNIF & 0.316703 & 0.000622\\
    \hline
  \end{tabular}
\end{table}
Based on these results, we conclude that:
\begin{enumerate}
  \item A-rated firm has a smaller Brownian motion part compared with
Ba-rated firm, besides B-rated firm has large $\mu_Z$ and especially
large $\sigma_Z$, which indicate that the loss due to sudden
economic hazard may fluctuate a lot for Ba-rated firms.
  \item The density function of A-rated firm still has the trend to
increase, which means the default rate of A-rated firm may increase
little faster in future. As for Ba-rated firm, its density functions
has decreased, so its default rate may increase very slowly or be
kept at a constant level.
  \item From the CPU time in Table \ref{Tab:hopt:CPU}, we can conclude that
the multivariate UNIF approach is much more efficient compared to
the conventional Monte-Carlo method.
\end{enumerate}

Our final example concerns with the default correlation of the two
firms. We use the following conditions in our multivariate UNIF
method:
\begin{enumerate}
  \item Setting $X(0)=2$ and $\ln(\kappa)=0$ for all firms.
  \item Setting $\gamma=\mu$ and $\mu=-0.001$ for all firms.
  \item Since we are considering two correlated firms, we choose $\sigma$ as,
\begin{equation}
  \sigma=\left[
  \begin{tabular}{cc}
    $\sigma_{11}$ & $\sigma_{12}$\\
    $\sigma_{21}$ & $\sigma_{22}$
  \end{tabular}\right],
\end{equation}
and
\begin{equation}
  \sigma\sigma^\top=\left[
  \begin{tabular}{cc}
    $\sigma_1^2$ & $\rho\sigma_1\sigma_2$ \\
    $\rho\sigma_1\sigma_2$ & $\sigma_2^2$
  \end{tabular}\right].
  \label{Eq:Brownian:corr}
\end{equation}
In Eq. (\ref{Eq:Brownian:corr}), $\rho$ reflects the correlation of
diffusion parts of the state vectors of the two firms, we set
$\rho=0.4$ as in \cite{Zhou:2001:corr}. Furthermore, we use the
optimized $\sigma_1$ and $\sigma_2$ in Table \ref{Table:param:one}
for A- and Ba-rated firms, respectively. Assuming $\sigma_{12}=0$,
we get,
\[
  \left\{
  \begin{tabular}{l}
    $\sigma_{11}=\sigma_1$,\\
    $\sigma_{12}=0$,\\
    $\sigma_{21}=\rho_{12}\sigma_2$,\\
    $\sigma_{22}=\sqrt{1-\rho_{12}^2}\sigma_2$.
  \end{tabular}\right.
\]
  \item The arrival rate for jumps satisfies the Poisson distribution with
intensity parameter $\lambda=0.1$ for both firms. The jump size is a
normal distribution $Z_t\sim N(\mu_{Z_i},\sigma_{Z_i})$, where
$\mu_{Z_i}$ and $\sigma_{Z_i}$ can be different for different firms
to reflect specifics of the jump process for each firm. We adopt the
optimized parameters given in Table \ref{Table:param:one}.
  \item As before, we generate the same interjump times $(T_j-T_{j-1})$ that
satisfy an exponential distribution with mean value equals to 1.
\end{enumerate}

The simulated default correlations can be obtained via the following
formula:
\begin{equation}
  \rho_{12}=\frac{1}{N}\sum_{n=1}^{N}
  \frac{P_{12,n}-P_{1,n}P_{2,n}}{\sqrt{P_{1,n}(1-P_{1,n})P_{2,n}(1-P_{2,n})}},
  \label{Eq:simulate:corr}
\end{equation}
where $P_{12,n}$ is the probability of joint default for firms 1 and
2 in each Monte Carlo cycle, $P_{1,n}$ and $P_{2,n}$ are the
cumulative default rates of firm 1 and 2, respectively, in each
Monte Carlo cycle.

\begin{figure}[hbtp]
  \centering
  \includegraphics[width=6.5cm]{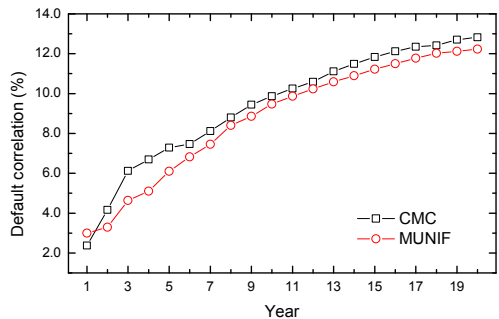}
  \caption{Default correlation (\%) of A- and Ba-rated firms. The
simulations were performed with Monte-Carlo runs $N=100,000$, for
the conventional Monte-Carlo method, the discretization size of time
horizon was $\Delta=0.005$.}
  \label{Fig:correlation}
\end{figure}

The simulated default correlations of A- and Ba-rate firms are given
in Fig. \ref{Fig:correlation}. Based on these results, we can
conclude that
\begin{enumerate}
  \item The default correlations tend to increase over long horizons and may
converge to a stable value.
  \item Our developed methodology gives almost identical default correlations
compared with the conventional Monte-Carlo method which confirms the
validity of the developed methodology.
\end{enumerate}

\section{Conclusions}
\label{conclusion}

\noindent In this contribution, we developed efficient
Monte-Carlo-based computational procedures for the solution of the
FPT problem in the context of multivariate (and correlated)
jump-diffusion processes. This was achieved by combining a fast
Monte-Carlo method for one-dimensional jump-diffusion process and
the generation of correlated multidimensional variables. The
developed procedures were applied to the analysis of multivariate
and correlated jump-diffusion processes. We have also discussed the
implementation of the developed Monte-Carlo-based technique for
multivariate jump-diffusion processes driving by several compound
Poisson shocks. Finally, we have applied the developed technique to
analyze the default events of two correlated firms via a set of
historical default data. The developed methodology provides an
efficient computational technique that is applicable in other areas
of credit risk and pricing options.

\section{Acknowledgements}

\noindent This work was supported by NSERC.



\footnotesize

\begin{thebibliography}{99}

\bibitem{Metwally:2002}
  S. Metwally and A. Atiya,
  Using Brownian bridge for fast simulation of jump-diffusion processes and barrier options,
  \textit{J. Derivatives}, \textbf{10}(2002), 43--54.

\bibitem{Zhang:1997}
  X. Zhang,
  Numerical analysis of American option pricing in a jump-diffusion model,
  \textit{Math. Oper. Res.}, \textbf{22}(1997), 668--690.

\bibitem{Zhou:2001:jump}
  C. Zhou,
  The Term Structure of Credit Spreads with Jump Risk,
  \textit{J. of Bank. Financ.}, \textbf{25}(2001), 2015--2040.

\bibitem{Atiya:2005}
  A. F. Atiya and S. A. K. Metwally,
  Efficient Estimation of First Passage Time Density Function for Jump-Diffusion Processes,
  \textit{SIAM J. Sci. Comput.}, \textbf{26}(2005), 1760--1775.

\bibitem{Cont:2003}
  R. Cont and P. Tankov,
  Financial Modelling with Jump Processes,
  Chapman \& Hall/CRC Press, London, 2003.

\bibitem{Crescenzo:2005}
  A. D. Crescenzo, E. D. Nardo and L. M. Ricciardi,
  Simulation of First-Passage Times for Alternating Brownian Motions,
  \textit{Methodol. Comput. Appl. Probab.}, \textbf{7}(2005), 161--181.

\bibitem{Black-Cox:1976}
  F. Black and J. C. Cox,
  Valuing corporate securities: Some effects of bond indenture provisions,
  \textit{J. Financ.}, \textbf{31}(1976), 351--367.

\bibitem{Zhou:2001:corr}
  C. Zhou,
  An analysis of default correlation and multiple defaults,
  \textit{Rev. Finan. Stud.}, \textbf{14}(2001), 555--576.

\bibitem{Hull:2001}
  J. Hull and A. White,
  Valuing Credit Default Swaps II: Modeling Default Correlations,
  \textit{J. Derivatives}, \textbf{8}(2001), 12--22.

\bibitem{Silverman:1986}
  B. W. Silverman,
  Density Estimation for Statistics and Data Analysis,
  Chapman and Hall, New York, 1986.

\bibitem{Glasserman:2004}
  P. Glasserman,
  Monte Carlo Methods in Financial Engineering,
  Springer, New York, 2004.

\bibitem{Platen:2003}
  P. E. Kloeden, E. Platen and H. Schurz,
  Numerical Solution of SDE Through Computer Experiments,
  Springer, Germany, 2003.

\bibitem{Chen:2005}
  J. T. Chen,
  Using the sum-of-uniforms method to generate correlated random variates with certain marginal distribution,
  \textit{Eur. J. Oper. Res.}, \textbf{167}(2005), 226--242.


\bibitem{Zhang:2006a}
  D. Zhang and R. V. N. Melnik,
  Efficient estimation of default correlation for multivariate jump-diffusion processes,
  \textit{Financ. Stoch.}, submitted.

\bibitem{Zhang:2006b}
  D. Zhang and R. V. N. Melnik,
  Monte-Carlo Simulations of the First Passage Time for Multivariate Jump-Diffusion Processes in Financial Applications,
  \textit{Quant. Financ.}, submitted.

\end{thebibliography}
\end{document}